\documentclass[aip,graphicx]{revtex4-1}

\usepackage{graphicx}
\usepackage{dcolumn}
\usepackage{amsmath,amsfonts}

\begin{document}

\preprint{}

\title{Quasi-total absorption peak by use of a backed rigid frame porous layer with circular periodic inclusions embedded} 



\author{J.-P. Groby}
\email[]{Jean-Philippe.Groby@univ-lemans.fr}
\affiliation{Laboratoire d'Acoustique de l'Universit\'e du Maine, UMR6613 CNRS/Univ. du Maine, Avenue Olivier Messiaen, F-72085 Le Mans Cedex 9, France.}
\author{O. Dazel}
\affiliation{Laboratoire d'Acoustique de l'Universit\'e du Maine, UMR6613 CNRS/Univ. du Maine, Avenue Olivier Messiaen, F-72085 Le Mans Cedex 9, France.}
\author{A. Duclos}
\affiliation{Laboratoire d'Acoustique de l'Universit\'e du Maine, UMR6613 CNRS/Univ. du Maine, Avenue Olivier Messiaen, F-72085 Le Mans Cedex 9, France.}
\author{L. Boeckx}
\affiliation{Huntsman Europe, Everslaan 45, B-3078 Everberg, Belgium.}
\author{W. Lauriks}
\affiliation{Laboratory of  Acoustic and Thermal Physics, KULeuven, Celestijnenlaan 200-D, B-3001 Heverlee, Belgium.}


\date{\today}

\begin{abstract}
The acoustic properties of a porous sheet of medium resistivity backed by a rigid plate in which are embedded a periodic set of circular inclusions is investigated. Such a structure behaves like a multi-component diffraction gratings. Numerical results show that this structure presents a quasi-total absorption peak below the essential spectrum, i.e. below the frequence of the fundamental quarter-wavelength resonance of the porous sheet in absence of inclusions. This result is explained either by the excitation of a complex trapped mode, or by the increase of viscous loss associated with a larger velocity gradient inside the layer at the modified quarter-wavelength resonance frequency. When more than one inclusion per spatial period are considered, additional quasi-total frequency peak are observed. The numerical results as calculated with the help of the mode-matching method described in the paper agree those as calculated with the help of a Finite Element method.
\end{abstract}


\maketitle 

\section{Introduction}\label{section1}
This work was initially motivated by a design problem connected to the determination of the optimal profile of discontinuous spatial distribution of porous materials and geometric properties for the absorption of sound. Porous materials (foams) suffer from a lack of absorption at low frequencies, when compared to their efficiency at higher frequencies. The purpose of the present article is to inverstigate an alternative to multi-layering by considering periodic inclusions embedded in a porous sheet attached on a rigid plate. This configuration results in a diffraction grating and possibly a sonic-crystal used in reflection.

The influence of a volumic heterogeneity addition on absorption and transmission of a porous layer was previously investigated by use of the multipole method in\cite{groby_wrcm2008,groby_jasa2010}, by embedding a periodic set of high-contrast inclusions, whose sizes are comparable to the wavelength, in a macroscopically-homogeneous porous layer whose thickness and weight are relatively small. This leads either to an increase of the absorption coefficient in the case of one layer of inclusions or to band-gaps and a total absorption peak in case of multi-layered set of inclusions (sonic crystal). The influence on the absorption was explained by mode excitation of the configuration enabled by the periodic inclusions, whose structure leads to energy entrapment. Other works related to volumic heterogeneities in macroscopically homogeneous porous material were carried out essentially by means of the homogeneization procedure\cite{tournat,gourdon2010}, and possibly leading to double porosity materials\cite{olny}.

The influence of the irregularities of the rigid plate on which are often attached porous sheets on the absorption coefficient was previsouly investigated by use of the multi-modal method in\cite{groby_jasa2010}, by considering periodic rectangular irregularities filled with air. This particularly leads in the case of one irregularity per spatial period to a total absorption peak associated with excitation of the fundamental modified mode of the backed plate. The latter is excited thanks to the surface grating. Other works related to surface irregularities were carried out, notably related to local resonances associated with fractal irregularities\cite{sapoval1,sapoval}.

Local resonance and trapped modes are an other possibility to localize the field. Trapped mode were largely studied in guided waves\cite{LintonWM2007} or in periodic structures\cite{porter2005}. Here, we investigate theoretically and numerically the influence on the absorption coefficient of the embedding of multiple inclusions grating in a rigid frame porous layer glued against a rigid wall. The effects of the modified mode of the plate and Bragg interference are clearly visible on the absorption curve, while a quasi-total absorption is obtained for a frequency below the one of the fundamental quarter-wavelength resonance of the backed porous sheet. The latter peak presents some of the specific features of a trapped mode excitation.  
\section{Formulation of the problem}\label{section2}
\subsection{Description of the configuration}\label{section2ss1}
\begin{figure}
\includegraphics[width=8.0cm]{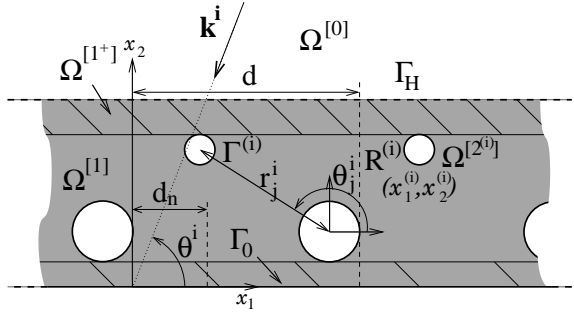}
\caption{Cross-sectional plane view of the configuration.}
\label{section2s1fig1}
\end{figure}

Both incident plane acoustic wave and plate are assumed to be
invariant with respect to the Cartesian coordinate $x_{3}$. A
cross-sectional $x_{1}\!-\!x_{2}$ plane view of the 2D scattering problem is
shown in figure \ref{section2s1fig1}.

Before the addition of the cylindric inclusions, the layer is made of a porous material saturated by air (e.g., a foam) which is modeled (by homogenization) as an (macroscopically homogeneous) equivalent fluid $M^{[1]}$. The porous sheet is backed by a rigid surface. The upper and lower flat and mutually-parallel boundaries of the layer, whose $x_2$ coordinates are $H$ and $0$, are designated by $\Gamma_H$ and $\Gamma_0$ respectively. $M^{[0]}$ and $M^{[1]}$ are in firm contact through $\Gamma_{H}$, i.e. the pressure and normal velocity are continuous across $\Gamma_{H}$ ($[p(\mathbf{x})]=0$ and $[\rho ^{-1} \partial_n p(\mathbf{x})]=0$, wherein $\mathbf{n}$ denotes the generic unit vector normal to a
boundary and $\partial _n$ designates the operator $\partial
_n=\mathbf{n} \cdot \nabla$). $\Gamma_0$ is rigid (Neumann type boundary conditions, $\partial_n p(\mathbf{x})=0$).

$N^c$ inclusions with a common spatial periodicity $d$ are embedded in the porous layer, that create a diffraction grating in the $x_1$ direction. Depending on the arrangement of the $N^{c}$ inclusions in the unit cell, a diffraction grating or a sonic-crystal of period $d^{c}$ can be formed ($d^{c}\leq d$). The set of indices by which the cylinders within the unit cell are identified is denoted by $\mathcal{N}^c\in\mathbb{N}$. The $j$-th inclusion occupies the disk $\Omega^{[2^{(j)}]}$ of radius $R^{(j)}$ and is centered at $\mathbf{x}^{(j)}=(x_1^{(j)},x_2^{(j)})$, $j\in\mathcal{N}^c$. The inclusions are infinitely rigid (Neumann type boundary conditions on $\Gamma^{(j)}$), i.e. the contrast between the elastic material $M^{[2]}$ and $M^{[1]}$ is very large. This also means that the inclusions can consist in tubes or holes posteriorly proofed for acoustic waves. Two subspaces $\Omega^{[1^{\pm}]}\in\Omega^{[1]}$ are also defined respectively corresponding to the upper and lower part of the plate not without inclusions.

The total pressure, wavenumber and wave speed are denoted by the generic
symbols $p$, $k$ and $c$ respectively, with $p=p^{[0]},~k=k^{[0]}=\omega/c^{[0]}$ in
$\Omega^{[0]}$ and $p=p^{[1]},~k=k^{[1]}=\omega/c^{[1]}$ in $\Omega^{[1]}$.

Rather than to solve directly for the  pressure $\bar{p}(\mathbf{x},t)$
(with $\mathbf{x}=(x_{1},x_{2})$), we prefer to deal with
$p(\mathbf{x},\omega)$, related to $\bar{p}(\mathbf{x},t)$ by the Fourier
transform $\displaystyle \bar{p}(\mathbf{x},t)=\int_{-\infty}^{\infty}p(\mathbf{x},\omega)e^{-\mbox{i}\omega
t} d\omega$. Henceforth, we drop the $\omega$ in $p(\mathbf{x},\omega)$ so as to denote the latter by $p(\mathbf{x})$.

The wavevector $\mathbf{k^{i}}$ of the incident plane wave lies in the sagittal plane and the angle of incidence is $\theta^{i}$ measured counterclockwise from the positive $x_{1}$ axis. The incident wave propagates initially in $\Omega^{[0]}$ and is expressed by $\displaystyle p^{i}(\mathbf{x})=A^{i}e^{\mbox{i}(k_{1}^{i}x_{1}-k_{2}^{[0]i}(x_{2}-H))}$, wherein $k_1^i=-k^{[0]}\cos \theta^i$, $k_2^{[0]i}=k^{[0]}\sin \theta^i$ and $A^{i}=A^{i}(\omega)$ is the signal spectrum.

The plane wave nature of the incident wave and the periodic nature of $\displaystyle \bigcup_{j\in\mathcal{N}^c}\Omega^{[2^{(j)}]}$ imply the Floquet relation
\begin{equation}
p(x_1+qd,x_2)=p(x_1,x_2)e^{\mbox{i}k_1^i qd}\,;\,\forall \mathbf{x}
\in \mathbb{R}^{2}\,;\,\forall q \in \mathbb{Z}~. \label{section2ss1e1}
\end{equation}

Consequently, it suffices to examine the field in the central cell of the plate.

The uniqueness of the solution to the  forward-scattering problem is ensured by the radiation condition:
\begin{equation}
p^{[0]}(\mathbf{x})-p^{i}(\mathbf{x})\! \sim\! \mbox{outgoing waves;
}\left|\mathbf{x} \right|\rightarrow \infty,~x_2\!>\!H~. \label{section2ss1e2}
\end{equation}
\subsection{Material modeling}\label{section2ss2}
Rigid frame porous material $M$ is modeled using the Johnson-Champoux-Allard model. The compressibilty $K$ and density $\rho$, linked to the sound speed through $c=\sqrt{1/\left(K\rho\right)}$ are\cite{Allard,deryck2007}
\begin{equation}
\begin{array}{ll}
\displaystyle \frac{1}{K}=&\displaystyle\frac{\gamma P_0}{\displaystyle \phi\left(\gamma-\left(\gamma-1 \right)\left(1+\mbox{i}\frac{\omega_c'}{\mbox{Pr}\omega}G(\mbox{Pr}\omega) \right)^{-1} \right)}~,\\[24pt]
\displaystyle \rho=&\displaystyle\frac{\rho_f \alpha_{\infty}}{\phi}\left(1+\mbox{i}\frac{\omega_c}{\omega}F(\omega) \right)~,
\end{array}
\label{section2ss2e1}
\end{equation}
wherein $\displaystyle \omega_c=\sigma \phi/\rho_f \alpha_{\infty}$ is the Biot frequency, $\displaystyle \omega_c'=\sigma' \phi/\rho_f \alpha_{\infty}$, $\gamma$ the specific heat ratio, $P_0$ the atmospheric pressure, $\mbox{Pr}$ the Prandtl number, $\rho_f$ the density of the fluid in the (interconnected) pores, $\phi$ the porosity, $\alpha_\infty$ the tortuosity, $\sigma$ the flow resistivity, and $\sigma'$ the thermal resistivity. The correction functions $G(\mbox{Pr}\omega)$ \cite{allardchampoux} and $F(\omega)$ \cite{johnson} are given by
\begin{equation}
\begin{array}{ll}
\displaystyle G(\mbox{Pr}\omega)=&\displaystyle \sqrt{1-\mbox{i} \eta \rho_f \mbox{Pr}\omega \left(\frac{ 2\alpha_{\infty}}{\sigma' \phi \Lambda'}\right)^2} ~,\\[8pt]
\displaystyle F(\omega)=&\displaystyle \sqrt{1-\mbox{i} \eta \rho_f \omega \left(\frac{ 2\alpha_{\infty}}{\sigma \phi \Lambda}\right)^2    }~,
\end{array}
\label{section2ss2e2}
\end{equation}
where $\eta$ is the viscosity of the fluid, $\Lambda'$ the thermal characteristic length, and $\Lambda$ the viscous characteristic length. The ``thermal resistivity'' is related to the thermal characteristic length through $\sigma'=8 \alpha_\infty \eta/\phi \Lambda'^2$, \cite{allardchampoux}.

The configuration is similar to but also more complex than those already studied in\cite{groby_wrcm2008,groby_jasa2009}, in sense the porous sheet is backed by a rigid plate, but also in sense the unit cell can be composed of more than one non overlapping inclusion, when the $x_2$-coordinates of the center of two inclusions are separated by a distance lower than the sum of their radii. In this latter case, the interaction between these inclusions can not be modeled as exposed in\cite{groby_wrcm2008} and a more complex interaction model should be employed\cite{botten_2}. The method of solution is also briefly summarized hereafter. 
\subsection{Field representations in $\Omega^{[0]}$ and $\Omega^{[1^{\pm}]}$}\label{section2ss3}
The continuity relations across the interfaces $\Gamma_H$ and $\Gamma_0$ are first considered in section \ref{section3ss1}. The field representations in $\Omega^{[0]}$, and $\Omega^{[1^{\pm}]}$ are needed as the first step. The continuity conditions
across $\Gamma^{(j)}$, $\forall j \in \mathcal{N}^c$ will be treated in section \ref{section3ss2}.

Separation of variables, the radiation condition, and the Floquet
theorem lead to the representation:
\begin{equation}
p^{[0]}(\mathbf{x})=\sum_{q\in\mathbb{Z}}\left[e^{-\mbox{i}k_{2q}^{[0]}\left(x_2-H\right)}
\delta_{q}+R_q
e^{\mbox{i}k_{2q}^{[0]}\left(x_2-H\right)}\right]e^{\mbox{i}k_{1q}x_1}
\mbox{, }\forall \mathbf{x}\in \Omega^{[0]}~, \label{section2ss3e1}
\end{equation}
wherein $\delta_{q}$ is the Kronecker symbol, $\displaystyle
k_{1q}=k_1^i+2q\pi/d$,
$\displaystyle k_{2q}^{[0]}=\sqrt{(k^{[0]})^2-(k_{1q})^2}$, with
$\textrm{Re}\left(k_{2q}^{[0]}\right)\geq 0$ and $\textrm{Im}\left(k_{2q}^{[0]}\right)\geq
0$. $R_q$ is the reflection coefficient of the plane wave denoted by the subscript $q$.

It is first convenient to use Cartesian coordinates $\left(x_1,x_2\right)$ to represent the field in $\Omega^{[1^{\pm}]}$. The latter are composed of the diffracted field in the plate and the fields scattered  by the inclusions, whose form depends on the position of $\mathbf{x}$, either below or above the inclusions \cite{botten}. Refering to\cite{groby_wrcm2008}, whatever the arrangement of the inclusions, $x_2$ is always larger than $\textrm{max}_{j\in\mathcal{N}^{c}} \left(x_2^{(j)}+R^{(j)}\right)$ in $\Omega^{[1^{+}]}$, while $x_2$ is always smaller than $\textrm{min}_{j\in\mathcal{N}^{c}} \left(x_2^{(j)}-R^{(j)}\right)$ in $\Omega^{[1^{-}]}$. The total field in $\Omega^{[1^{\pm}]}$ can be written in Cartesian coordinates as
\begin{multline}
p^{[1^\pm]}(\mathbf{x})=\sum_{q\in\mathbb{Z}}\left(f_{q}
e^{-\mbox{i}k_{2q}^{[1]}x_2}+g_{q}
e^{\mbox{i}k_{2q}^{[1]}x_2}\right)e^{\mbox{i}k_{1q} x_1}\\
+\!\sum_{q\in\mathbb{Z}}\!\sum_{j\in\mathcal{N}^c}\!\sum_{l\in\mathbb{Z}} K_{ql}^{\pm} B_l^{(j)}
e^{\mbox{i}\left(k_{1q}\left(x_{1}-x_1^{(j)}\right)\pm k_{2q}^{[1]} \left(x_{2}-x_2^{(j)}\right)\right)}, \label{section2ss3e3}
\end{multline}
wherein $B_l^{(j)}$ are the coefficients of the field scattered  by the $j$-th cylinder of the unit cell, $f_q$ and $g_q$ are the coefficients of the diffracted waves inside the layer associated with the plane wave denoted by $q$, and $\displaystyle K_{ql}^{\pm}=2(-\mbox{i})^l e^{\pm\mbox{i}l\theta_q}/dk_{2q}^{[1]}
$ with $\theta_q$ such that $k^{[1]} e^{\mbox{i}\theta_q}=k_{1q}+\mbox{i}k_{2q}^{[1]}$, \cite{botten_2}.
\section{Determination of the acoustic properties of the configuration}\label{section3}
\subsection{Application of the continuity conditions across $\Gamma_H$ and $\Gamma_0$}\label{section3ss1}
Applying the continuity of the pressure field and of the normal component of the velocity across $\Gamma_H$ and the Neumann condition on $\Gamma_0$, introducing the proper field representation therein, Eqs.(\ref{section2ss3e1}) and (\ref{section2ss3e3}), and making use of the orthogonality relation $\displaystyle \int_{-\frac{d}{2}}^{\frac{d}{2}}
e^{\mbox{i}\left(k_{1n}-k_{1l}\right)x_1}\textrm{d}x_1= d \delta_{nl}$, $\forall (l,n) \in \mathbb{Z}^2$ give rise to a linear set of equations. After some algebra and rearrangements, this linear set reduces to a coupled system of equations for solution of $R_q$, $f_q$ and $g_q$ in terms of $B_l^{(j)}$.
\subsection{Application of the multipole method}\label{section3ss2}
The expressions of $f_q$ and $g_q$ in terms of $B_l^{(j)}$ are introduced in the so-denoted diffracted field inside the layer. The latter field accounts for the direct diffracted waves inside the layer and for the reflected waves at the boundaries $\Gamma_0$ and $\Gamma_H$ previously scattered by each inclusions. This expression when compared with the expression of the direct scattered field by the inclusions, is valid in the whole domain $\Omega^{[1]}$. To proceed further, the
Cartesian form of this field is converted to the cylindrical harmonic form in the polar coordinate system attached to each inclusion, as stated for example in\cite{groby_wrcm2008}. Effectively, central to the multipole method are the local field expansion or multipole expansions around each inclusion.

The pressure field in the vicinity of the $J$-th inclusion in the polar coordinate system attached to this inclusion and introducing $A_L^{(J)}$ the coefficient of the locally-incident field, reads as
\begin{multline}
p^{[1]}(\mathbf{r}_J)=\sum_{L\in\mathbb{Z}} B_L^{(J)}
\textrm{H}_L^{(1)}\left(k^{[1]}
r_J\right)e^{\mbox{i}L\theta_J}\\
+\sum_{L\in\mathbb{Z}}
\left[\sum_{l\in\mathbb{Z}} S_{L-l}B_l^{(J)} +\sum_{j\neq J}\sum_{l\in\mathbb{Z}} S_{L-l}^{(J,j)}B_l^{(j)}+\sum_{j\in\mathcal{N}^c}\sum_{l\in\mathbb{Z}}\sum_{q\in\mathbb{Z}}Q_{Llq}^{(J,j)}B_l^{(j)}+\sum_{q\in\mathbb{Z}} F_{qL}^{(J)}\right] \textrm{J}_L\left(k^{[1]}r_J\right)e^{\mbox{i}L\theta_J}\\[12pt]
=\sum_{L\in\mathbb{Z}} \left[B_L^{(J)}
\textrm{H}_L^{(1)}\!\left(k^{[1]}
r_J\right)+A_L^{(J)} \textrm{J}_L\!\left(k^{[1]}r_J\right)\right]e^{\mbox{i}L\theta_J},
\label{section3ss2e1}
\end{multline}
with
\begin{widetext}
\begin{equation}
\begin{array}{ll}
\displaystyle F_{qL}^{(J)}=&\displaystyle\frac{2\delta_q\alpha_q^{[0]}}{D_q}\cos \left( k_{2q}^{[1]}x_2^{(J)}-L\theta_q\right) e^{\mbox{i}k_{1q}^{[1]}x_1^{(J)}}~,\\[12pt]
\displaystyle S_{L-l}=&\displaystyle\sum_{i=1}^{\infty} \textrm{H}_{L-l}^{(1)}\left(k^{[1]} id \right)
\left[e^{\mbox{i}k_1^i id}+\left(-1 \right)^{L-l} e^{-\mbox{i}k_1^i id} \right]~,\\[12pt]
\displaystyle Q_{Llq}^{(J,j)}=&\displaystyle\frac{2(-\mbox{i})^{l-L}e^{\mbox{i}k_{1q}(x_1^{(J)}-x_1^{(j)})}}
{dk_{2p}^{[1]}D_p}\left[ \left(\alpha_q^{[1]}-\alpha_q^{[0]}\right)e^{\mbox{i}k_{2q}^{[1]}H} \cos\left(k_{2q}^{[1]} \left(x_2^{(j)}-x_2^{(J)}\right)-(l-L)\theta_q \right)\right.\\
\displaystyle &\displaystyle\left. +\alpha_q^{[1]} \cos\left(k_{2q}^{[1]} \left(x_2^{(j)}+x_2^{(J)}-H \right)-(l+L)\theta_q \right)+\mbox{i}\alpha_q^{[0]} \sin\left(k_{2q}^{[1]} \left(x_2^{(j)}+x_2^{(J)}-H \right)-(l+L)\theta_q \right)\right]~,\\[12pt]
\displaystyle D_q=&\displaystyle \alpha_q^{[0]}\cos\left(k_{2q}^{[1]}H \right)-\mbox{i}\alpha_q^{[1]}\sin\left(k_{2q}^{[1]}H \right)~,
\end{array}
\label{section3ss2e2}
\end{equation}
\end{widetext}
wherein $S_{L-l}$ is the lattice sum often refered to as Schl\"omilch series for non-dissipative material, $\textrm{H}_L^{(1)}$ is the $L$-th order Hankel function of first kind and $\textrm{J}_L$ is the $L$-th order Bessel function. The terms $S_{L-l}^{(J,j)}$ accounts for the coupling between the multiple inclusions inside the unit cell and take the form:
\begin{equation}
\displaystyle S_{Ll}^{(J,j)}=\sum_{q\in\mathbb{Z}}\frac{(-\mbox{i})^{L-l}2e^{\pm\mbox{i}(L-l)\theta_q}}{dk_{2q}^{[1]}}
 e^{\mbox{i}\left(k_{1q}(x_1^{(J)}-x_1^{(j)})\pm k_{2q}^{[1]}(x_2^{(J)}-x_2^{(j)}) \right)}\left(1-\delta_{Jj} \right)~,
\label{section3ss2e3}
\end{equation}
wherein the signs $+$ and $-$ correspond to $x_2^{J}\geq x_2^j$ and $x_2^{J}\leq x_2^j$ respectively, which can be found in\cite{groby_wrcm2008} when $|x_2^{(J)}-x_2^{(j)}|\geq R^{(j)}+R^{(i)}$ or
\begin{equation}
\displaystyle S_{Ll}^{(J,j)}=\left[B_l^{(j)} \mbox{H}_{L-l}^{(1)}\left(r_J^j\right)e^{\mbox{i}\left(l-L \right)\theta_J^j}\!+\!\sum_{o\in\mathbb{Z}}S_{o-l} B_l^{(j)}\mbox{J}_{L-o}\!\left(k^{[1]} r_J^j \right)e^{\mbox{i}\left(o-L \right)\theta_J^j}\right](1-\delta_{Jj})~,
\label{section3ss2e4}
\end{equation}
when $|x_2^{(J)}-x_2^{(j)}|\geq R^{(j)}+R^{(i)}$. This latter form agrees with the one found in\cite{botten_2} when the inclusions are aligned inside the unit cell, i.e. $x_2^{(j)}=x_2^{(J)}$, $\forall j \in \mathcal{N}^c$, which imposes $\theta_J^j=0$ or $\theta_J^j=\pi$. In Eq.(\ref{section3ss2e4}), $(r_J^j,\theta_J^j)$ is the coordinate of $\mathbf{x}^{(j)}$ in the polar coordinate system attached to the $J$-th inclusions, i.e. centered at $\mathbf{x}^{(J)}$.

Finally, it is well known that the coefficients of the scattered field and those of the locally-incident field are linked by a matrix relation derived from the boundary condition on $\Gamma^{(J)}$ only, i.e., $B_L^{(J)}=V_L^{(J)} A_L^{(J)}$, wherein $V_L^{(J)}$ are the cylindrical harmonic reflection coefficients. These coefficients takes the form $V_L^{(J)}=-\dot{\mbox{H}}_{L}^{(1)}\left(k^{[1]}R^{(j)}\right)/\dot{\mbox{J}}_{L}\left(k^{[1]}R^{(j})\right)$ in case of Neumann type boundary condition, with $\dot{\chi}(x)=d\chi/dx$, $\chi$ being either Hankel or Bessel functions. Introducing the expression of $A_L^{(J)}$ derived from Eq.(\ref{section3ss2e1}) in the previous relation gives rise to the linear system of equations for the solution of $B_L^{(j)}$. This linear system may be written in the matrix form, when denoting by $\mathbf{B}$ the infinite column matrix of components $B_L^{(j)}$
\begin{equation}
\left(\mathbf{I}-\mathbf{V}\left(\mathbf{S}+\mathbf{Q}\right)\right)\mathbf{B}=\mathbf{V}\mathbf{F}~,\\[8pt]
\label{section3ss2e5}
\end{equation}
wherein $\mathbf{F}$ is a vector of components $\displaystyle \sum_{q\in\mathbb{Z}} F_{qL}^{(J)}$, which accounts for the solicitation of the $J$-th inclusion by a wave that is previously diffracted inside the layer, $\mathbf{V}$ is a diagonal square matrix of components $V_L^{(J)}$, $\mathbf{S}$ and $\mathbf{Q}$ are two matrices of components $S_{L-l}+S_{L-l}^{(J,j)}$, which accounts for the coupling between the $J$-th and the $j$-th inclusion inside the layer and $\displaystyle \sum_{q\in\mathbb{Z}} Q_{Llq}^{(J,j)}$, which accounts for the coupling between the $J$-th and the $j$-th inclusion through waves diffracted by the layer.

The expressions of the components involved in $\left(\mathbf{I}-\mathbf{V}\left(\mathbf{S}+\mathbf{Q}\right)\right)\mathbf{B}=\mathbf{V}\mathbf{F}$ are identical to those found in\cite{groby_wrcm2008}, when the half-space behind the layer vanishes and when the center of the cylinders are defined as they are in the present article.
\subsection{Evaluation of the fields, reflection and absorption coefficients.}\label{section3ss3}

Once the linear system (\ref{section3ss2e5}) is solved, the expression of $R_p$ in terms of $B_l^{(j)}$ reads as
\begin{multline}
R_q=\frac{\alpha^{[0]i}\cos\left(k_{2}^{[1]i}L \right)+\mbox{i}\alpha^{[1]i}\sin\left(k_{2}^{[1]i}L \right)}{D^{i}}\\
+\sum_{q\in\mathbb{Z}}\sum_{j\in\mathcal{N}^c} \sum_{l\in \mathbb{Z}}\frac{4\left(-\mbox{i}\right)^{l}\alpha_q^{[1]}}{d k_{2q}^{[1]}D_p}B_l^{(j)}\cos\left(k_{2q}^{[1]}x_2^{(j)}-l\theta_q\right) e^{-\mbox{i}k_{1q}x_1^{(j)}}~.
\label{section3ss3e2}
\end{multline}

The first term corresponds to the reflection coefficient in terms of waves in absence of inclusion, i.e. for $q=0$ or for the incident plane component indexed by $i$, and the second terms accounts for the inclusions.

Introduced in Eq.(\ref{section2ss3e1}), the reflected field is expressed as a sum of the field in the absence of the inclusions with the field due to the inclusions.
In case of an incident plane wave with spectrum $A^{i}(\omega)$, the conservation of energy relation takes the form
\begin{equation}
1=\mathcal{R}+\mathcal{A}~,
\label{section3ss4e1}
\end{equation}
with $\mathcal{R}$ and $\mathcal{A}$ the hemispherical reflection and the absorption coefficients. $\mathcal{R}$ is defined by
\begin{equation}
\displaystyle \mathcal{R}=\sum_{q\in\mathbb{Z}}\frac{\mbox{Re}\left(k_{2q}^{[0]} \right)}{k_{2}^{[0]i}}\frac{\|R_q\|^2}{\|A^i \|^2 }=\displaystyle \sum_{q=-\widetilde{q}_{-}}^{\widetilde{q}_{+}}\frac{k_{2q}^{[0]}}{k_{2}^{[0]i}}\frac{\|R_q\|^2}{\|A^i \|^2 }~,
\label{section3ss4e2}
\end{equation}
 wherein $\widetilde{q}_{\mp}$ are such that $\widetilde{q}_{\mp} <d/2\pi\left(k^{[0]}\pm k_1^{i} \right)<\widetilde{q}_{\mp}+1$ and the expression of $R_p$ are given equation (\ref{section3ss3e2}). $\mathcal{A}$ takes the form
\begin{equation}
\displaystyle \mathcal{A}=\frac{1}{dk_2^{[0]i}\|A^i \|^2 }\left( \mathcal{A}_{D}+\mathcal{A}_{S}\right)~,
\label{section3ss4e3}
\end{equation}
whererin
\begin{equation}
\displaystyle\mathcal{A}_{D}=\frac{\rho^{[0]}}{\mbox{Re}\left(\rho^{[1]}\right)} \int_{\Omega_{[1]}}\displaystyle \mbox{Im}\left( \left( k^{[1]}\right)^{2}\right) \|p^{[1]}\left(\mathbf{x}\right) \|^{2} \textrm{d}\bar{\omega}~,
\label{section3ss4e4}
\end{equation}
is the inner absorption of domain $\Omega^{[1]}$. $\textrm{d}\bar{\omega}$ is the differential element of surface in the sagittal plane and 
\begin{equation}
\displaystyle\mathcal{A}_{S}\! =\!\mbox{Re} \int_{\Gamma_a} \frac{\rho^{[0]}}{\rho^{[1]}}�\frac{\mbox{Im} \left( \rho^{[1]}\right)}{\mbox{Re}\left(\rho^{[1]} \right)} p^{[1]\star}\left(\mathbf{x}\right)\mathbf{\nu}_{01} \cdot \nabla p^{[1]} \left(\mathbf{x}\right)\textrm{d} \gamma,
\label{section3ss4e5}
\end{equation}
is the surface absorption related to interfaces $\Gamma_H$. $\textrm{d}\gamma$ is the differential arc length in the cross-sectional plane, $\mathbf{\nu}_{01}$ is the outward-pointing unit vector to the boundary $\Gamma_H$, and $p^{*}$ is the complex conjuguate of $p$.

$\mathcal{A}_{S}$ accounts for the absorption induced by the viscous dissipation at the interfaces. Effectively, it is obvious from Eq.(\ref{section3ss4e5}) that $\mathcal{A}_{S}$ does not vanish because of non-vanishing $\mbox{Im}\left(\rho^{[1]}\right)$, which is a consequence of the modeling of viscous dissipation phenomenon\cite{zwickerkosten}.

Because of the complicated shape of $\Omega^{[1]}$ and the non-vanishing term $\mathcal{A}_{S}$, $\mathcal{A}$ will not be calculated by the expression given in (\ref{section3ss4e3}), but rather by $\mathcal{A}=1-\mathcal{R}$.
\section{Numerical results, validation and discussion}\label{section4}
The infinite sum $\sum_{q\in\mathbb{Z}}$ over the indices of the $k_{1q}$ depends on the frequency and on the period of the grating. An empirical truncation rule is employed, inspired from\cite{groby_wrcm2008,groby_jasa2009,groby_jasa2010} and determined by performing a large number of numerical experiments $\sum_{q=-Q_-}^{Q_+}$ such that $Q_\mp=\textrm{int}\left(d/2\pi\left(3\textrm{Re}\left(k^{[1]}\right)\pm k_{1}^i \right)\right) +10$. In the latter equations, $\textrm{int}\left(a\right)$ represents the integer part of $a$.

The infinite sum $\sum_{m\in\mathbb{Z}}$ over the indices of the modal representation of the
  diffracted field by a cylinder is truncated\cite{barber} as $\sum_{m=-M}^{M}$ such that $M=\mbox{int}\left(\mbox{Re}\left(4.05 \times \left(k^{[1]} R \right)^{\frac{1}{3}}+k^{[1]} R \right) \right)+10$.

Finally, the infinite sum (lattice sum) embedded in $S_{L-l}$ in equations (\ref{section3ss2e2}) and (\ref{section3ss2e4})  $\sum_{i=1}^{\infty}$ is found to be slowly convergent, particularly in the absence of dissipation, and is found to be strongly dependent on the indice $L-l$. A large literature exists on this  problem \cite{twersky,linton}. Here, the fact that the medium $M^{[1]}$ is dissipative greatly simplifies the evaluation of the Schl\"omilch series. The superscript $I$ in $S_{L-l}^{\{I\}}$ identifies the integer over which the sum is performed, i.e. $\sum_{i=1}^{I}$. This sum is carried out until the conditions $\left|\mbox{Re}\left(\left(S_{L-l}^{\{I+1\}}-S_{L-l}^{\{I\}}\right)/S_{L-l}^{\{I\}} \right)\right|\leq 10^{-5}$ and $\left|\mbox{Im}\left(\left(S_{L-l}^{\{I+1\}}-S_{L-l}^{\{I\}}\right)/S_{L-l}^{\{I\}} \right)\right|\leq 10^{-5}$ are reached \cite{groby_jasa2009}.
\begin{table}
\caption{Geometry of the considered configurations.}
	\begin{center}
		\begin{tabular}{c|c|c|c|c|c}
& N & $d$ (cm) & H (cm) & $(x_1^{(j)}~\textrm{(cm)},x_2^{(j)}~\textrm{(cm)})$ & $R^{(j)}$ (cm) \\ 
 \hline
C1 & 1 & 2 & 2 & (1,1) & 0.75 \\
C2 & 2 & 2 & 3.5 & $(x_1^{(1)},x_2^{(1)})=(1,1)$ & $R^{(1)}=0.75$ \\
 & & && $(x_1^{(2)},x_2^{(2)})=(0.5,1+\sqrt{3}d/2)$ & $R^{(2)}=0.5$ \\
		\end{tabular}
	\end{center}
\label{table1}
\end{table}

Numerical calculations have been performed for various geometrical parameters whose values are reported in Table \ref{table1}, and whithin the frequency range of audible sound, particularly at low frequencies. For all calculations, the ambient and saturating fluid is air ($\rho^{[0]}=\rho_f=1.213\textrm{ kg.m}^{-3}$, $c^{[0]}=\sqrt{\gamma P_0/\rho_f}$, with $P_0=1.01325 \times 10^{5}\textrm{ Pa}$, $\gamma=1.4$, and $\eta=1.839 \times 10^{-5}\textrm{ kg.m}^{-3}\textrm{.s}^{-1}$). One of the main constraints in designing acoustically absorbing materials are the size and weight of the configuration. A particular attention is paid on the dimension, i.e. thickness, and the frequencies of the absorption gain, which have to be commonly as small as possible. The absorption gain is defined by reference to the absorption of the same configuration without inclusion embedded. The initial configuration consists in a $2\textrm{ cm}$ thick porous sheet of Fireflex (Recticel, Belgium) backed by a rigid plate. The material characteristics are reported in Table \ref{table2} and were determined by use of traditional methods\cite{Allard}. Circular cylinders of $7.5\textrm{ mm}$ radius are embedded with a spatial periodicity of $2\textrm{ cm}$.
\begin{table}
\caption{Parameters of the porous foam used in the article.}
	\begin{center}
		\begin{tabular}{c|c|c|c|c|c}
$\phi$ & $\alpha_{\infty}$ & $\Lambda$ ($\mu$m) & $\Lambda'$ ($\mu$m) & $\sigma$ (Ns.m$^{-4}$) & $\nu_c$ (Hz)\\ 
 \hline
0.95  & 1.42  & 180 & 360& 8900 & 781\\
		\end{tabular}
	\end{center}
\label{table2}
\end{table}

\subsection{One inclusion per spatial period}\label{section4ss1}
\begin{figure}
\includegraphics[width=8.0cm]{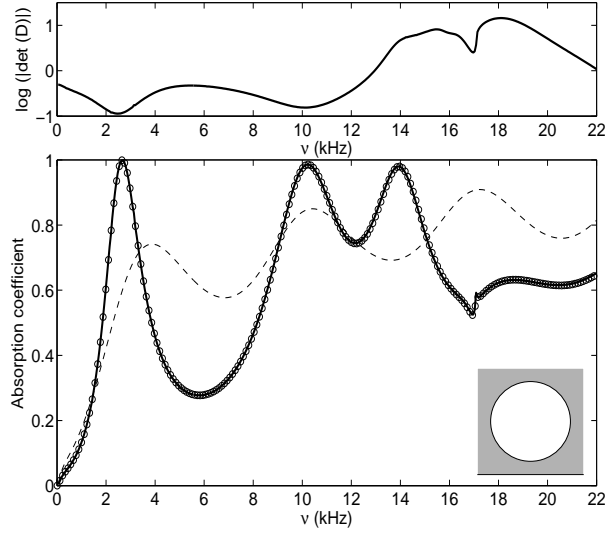}
\caption{Configuration C1 - Absorption coefficient of a $H=2\textrm{ cm}$ thick porous sheet of Fireflex backed by a rigid plate ($---$) without inclusion embedded and (---) with a $R=75\textrm{ mm}$ radius circular cylinder embedded per spatial period $d=2\textrm{ cm}$. The Finite Element result is plotted with (o). The absolute value of the determinant of the propagation matrix $\mathbf{D}=\mathbf{I}-\mathbf{V}\left(\mathbf{S}+\mathbf{Q}\right)$ is plotted on top of the figure.}
\label{section4s1fig2}
\end{figure}

We first consider only one inclusion centered in the unit cell, i.e. $(x_1^{(1)},x_2^{(1)})=(d/2,H/2)=(1\textrm{ cm},1\textrm{ cm})$.  The first two modified modes of the plate, which are excited because of the periodic arrangement of the inclusions, Appendix \ref{app1}, stand around $\nu_{(1,1)}\approx 14\textrm{ kHz}$ and $\nu_{(2,1)}\approx 16\textrm{ kHz}$. The attenuation associated with both modes is relatively large, figure\ref{section4s1fig2}.

\begin{figure}
\includegraphics[width=8.0cm]{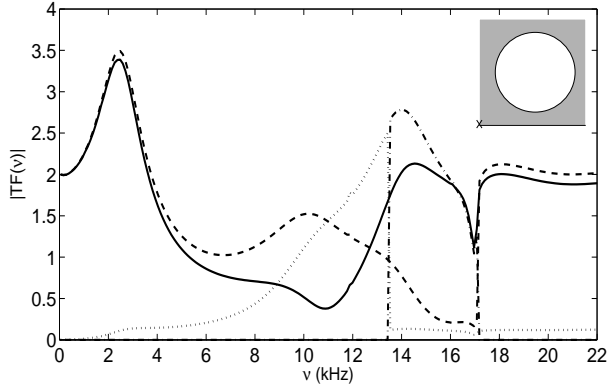}
\caption{Configuration C1 - Transfer function, TF (---), on $\Gamma_0$ at $1\textrm{ cm}$ from the center of the inclusion (between two inclusions), and its different contributions when the configuration is excited at normal incidence : ($--$) $TF_1$, ($-\cdot$) $TF_2$, and ($\cdot \cdot$) $TF_3$.}
\label{section4s1fig1}
\end{figure}

Different type of waves corresponds to each kind of mode related to the grating, i.e. mode of the grating (MG) and modified mode of the backed layer (MMBL): evanescent waves in $\Omega^{[1]}$ (and also in $\Omega^{[0]}$) for the MG, and evanescent waves in $\Omega^{[0]}$ and propagative waves in $\Omega^{[1]}$ for the MMBL. In order to determine which type of mode is excited by the plane incident wave, we have plotted in Figure \ref{section4s1fig1} the transfer function as calculated by $TF(\nu)=p(\mathbf{x},\omega)/p^{[0]i}(\mathbf{x},\omega)$ on $\Gamma_0$ ($x_2=0$) at $1\textrm{ cm}$ from the center of the inclusion (between two inclusions), when excited at normal incidence. The tranfert function is separated on the different intervals corresponding to the different type of waves that are involved in the total pressure calculation: $TF(\nu)$ is the total transfer function, $TF_1(\nu)$ is the contribution of the propagative waves in both $\Omega^{[0]}$ and $\Omega^{[1]}$, $TF_2(\nu)$ is the contribution of the evanescent waves in $\Omega^{[0]}$ and propagative ones in $\Omega^{[1]}$, and $TF_3(\nu)$ is the contribution of the evanescent waves in both $\Omega^{[0]}$ and $\Omega^{[1]}$. The transfer function possesses a large peak at $\approx 15\textrm{ kHz}$, around $\nu_{(1,1)}$ and $\nu_{(2,1)}$. This also proves that the MMBL are the most excited modes related to the grating, at least around these frequencies. The peak results from a continuous drop between evanescent waves in both material to evanescent waves in the air medium. This also means that this peak is neither a MMBL nor a MG, but result from a complex combination of these two types of mode, with a structure closer to the one of the MMBL. Because of this stucture, the energy is trapped in the layer leading to an increase in the absorption coefficient. The translation of the excitation of these modes in terms of absorption, i.e. the peak around $17\textrm{ kHz}$ figure \ref{section4s1fig2}, is smaller than the one depicted in \cite{groby_jasa2010}, because the attenuation associated with the modes in the present configuration is larger and because the flow resistivity of the foam considered here is larger. The design of a struture composed of a layer with inclusions embedded is more based on compromises than with irregularities of the rigid backing\cite{groby_jasa2010}. These compromises relate the spatial periodicity, the radius of the inclusion or better the ratio $R/d$ which can not be too small and which constrains the layer thickness and the properties of the latter.

Figure \ref{section4s1fig2} depicts the absorption coefficient as calculated for this geometry. This result was validated numerically by matching the absorption coefficient as calculated with the present method with the one as calculated with a Finite Element method. Quadratic finite element were used to approximate the pressure inside the unit cell, thereby leading to a discretized problem of 2196 elements and 1238 nodes. The periodicity relation, i.e. the Floquet condition, were applied on both sides of the discretized domain, i.e. at each nodes of $x_1$-coordinate $0$ and $d$. For this periodicity relation to be correctly implemented these two sides were discretized with similar nodes, i.e. idential $x_2$-coordinate. The results match well, thus validating the described method, figure \ref{section4s1fig2}.

Because of the rigid backing, which acts as a perfect mirror, the response of the configuration possesses some particular features related to multi-layered grating. We also introduce $d_2=2 x_2^{(1)}$ the distance between the center of the circular cylinder and the center of its image. Each grating is going to interfere with one another at the Bragg frequencies $\nu_{(n)}^{b}=n\pi \textrm{Re}\left(c^{[1]}\right)/d_2$. In particular, the first Bragg frequency is $\nu_{(1)}^{b}\approx 6 \textrm{ kHz}$. The latter frequency is largely employed to determine the central frequency of the band gaps for phononic crystals and corresponds to the maximum of reflected energy and so to a minimum of transmitted energy -band gaps- in case of phononic crystal. The absorption coefficient also presents a minimum at $\nu_{(1)}^{b}$.

A particular feature of the response of this configuration is that the absorption coefficient presents a peak close to unity at a low frequency $\nu_{t}$ below the so-called fundamental quarter-wave resonance frequency, i.e. below what can be associated with an essential spectrum. 

A sensitivity analysis, performed by varying one parameter while the other are kept constant at their value, shows that the radius of the inclusion has a large influence on $\nu_{t}$ and on the amplitude of the corresponding absorption peak. The radius $R$ was varied from $1\textrm{ mm}$ to $9.5\textrm{ mm}$. In terms of amplitude of the absorption peak, $R=75\textrm{ mm}$ is the optimal value, while $\nu_{t}$ decreases when $R$ increases. In the opposite, $d=2\textrm{ mm}$ is the optimal value in terms of the amplitude of the peak, but $\nu_{t}$ increases when $d$ increases from $1.75\textrm{ cm}$ to $3.75\textrm{ cm}$. The spatial periodicity of the arrangement acts inversely on $\nu_{t}$ than it does on the frequencies of the modes closely related to $d$ like the MMBL and the MG, Appendix \ref{app1}. In fact, the amplitude of the absorption peak increases with the filling ration $R/d$ until being close to unity and drastically decreases after this value because the wave can no more propagate in the layer towards the rigid backing and is mainly reflected on the cirular grating, or because the density of inclusion becomes insignificant. In the same way, when the angle of incidence decreases $[\pi/2;\pi/6]$, the $\nu_{t}$ increases in the opposite to the frequencies of the MMBL and of the MG. The amplitude of the peak is quite close to unity until $\pi/6$. For smaller values of incidence angle, the amplitude of this low frequency peak begins to decrease.

When $R=75\textrm{ mm}$ and $d=20\textrm{ mm}$, $\nu_t$ is all the smaller that the inclusion is distant from the rigid backing, i.e. that $x_2^{(1)}$, the center of the inclusion, is large. The amplitude of the peak is close to unity whatever $x_2^{(1)}$, in $[0.8\textrm{ cm} ;1.2\textrm{ cm}]$ the range conditionned by the layer thickness and the inclusion radius. 

\begin{figure}
\includegraphics[width=8.0cm]{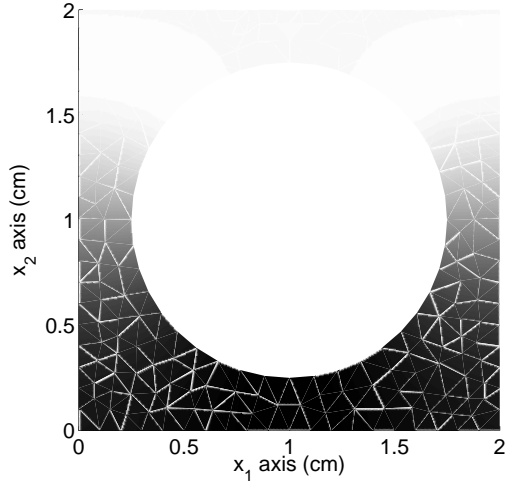}
\caption{Configuration C1 - Snapshot of the module of the pressure field inside the porous sheet at $\nu_t=2674\textrm{ Hz}$.}
\label{section4s1fig2b}
\end{figure}

The features of this low frequencies absorption peak sounds like phenomena that are related to trapped modes in wave guides\cite{LintonWM2007} or embedded Rayleigh-Bloch waves\cite{porter1999,porter2005}. These modes has finite energy and corresponds to a solution which decays down away from the perturbation. Figure\ref{section4s1fig2b} shows a snapshot of the module of the pressure field at $\nu_t$. The latter clearly exibits a maximum on the side of the rigid plate and a minimum on the side of $\Gamma_0$, which is typical of a trapped mode. Everything seems to happen like if a Dirichlet wave guide of  thickness $2H$ presents, necessarily, symetric obstacles formed by the inclusion and its image. In our case, these trapped modes are complex, because the boundaries of the wave-guide are not Dirichlet conditions but continuity conditions, and because $M^{[1]}$ is a dissipative medium. The determinant as calculated for the configuration C1, figure \ref{section4s1fig2}, present a minima at $\nu_{t}$, which suggests that a complex (trapped) mode CTM stands at this frequency. In the opposite, it is clear from figure\ref{section4s1fig1}, that the peak around $\nu_{t}$ is mainly associated with propagative waves in both domains, a small part of it being associated with evanescent waves in the layer which entraps the energy. This phenomena was already encountered in\cite{craster2009} and attributed to the periodicty of the configuration. An other explanation of the quasi-absorption peak is related to the modification of wave path and structure global properties inside the porous sheet. For a particular ratio $R/d$, the pressure gradient, which is a cause of viscous loss, is larger inbetween the inclusions and between the inclusions and the rigid backing. This loss phenomenon results from a continuous modification of the quarter-wavelength resonance when the inclusion radius increases.

Other layer thicknesses were tested. It was found that for each layer thickness, in the suitable range for the application, and a centered $(x_1^{(1)},x_2^{(1)})=(d/2,H/2)$ inclusion, a couple $(R^{(1)},d)$ exists for which a quasi-total absorption peak exists below the quarter-wavelength resonance frequency. 

Finally, because the parameters of a foam are often difficult to predict before its polymerization, a sensitivity analysis has been performed with regards to the acoustic and structural parameters of the porous sheet. The amplitude and frequency of the CTM is quasi independant from a variation of $\phi$, $\Lambda$ and $\Lambda'$. When $\alpha$ increases from $1.02$ to $1.42$, the sound speed in the material decreases and $\nu_{t}$ decreases, while the amplitude of the associated peak stands close to one. The resisgtivity $\sigma$ particularly influences the amplitude of the peak. When it increases, the amplitude admits a maximum and the peak is wider, while $\nu_{t}$ increasse. The resistivity $\sigma$ ($[3900\textrm{ Ns.m}^{-4};12900\textrm{ Ns.m}^{-4}]$) is the parameter that mostly influences the results and its value has to be close to the one used in the simulations, i.e. in our case, a value inbetween $7000\textrm{ Ns.m}^{-4}$ to $11000\textrm{ Ns.m}^{-4}$ is acceptable.


\subsection{Two or more inclusions per spatial period}\label{section4ss3}

Various configurations were tested, involving two or more inclusions per spatial period. The frequency band investigated stands below the quarter-wavelength resonance frequency of the associated porous sheet or at least below the first Bragg frequency, i.e. below the frequency of the first modified mode of the porous sheet. The lowest frequency boundary is naturally the solid-fluid decoupling frequency or at least the Biot frequency.

\begin{figure}
\includegraphics[width=8.0cm]{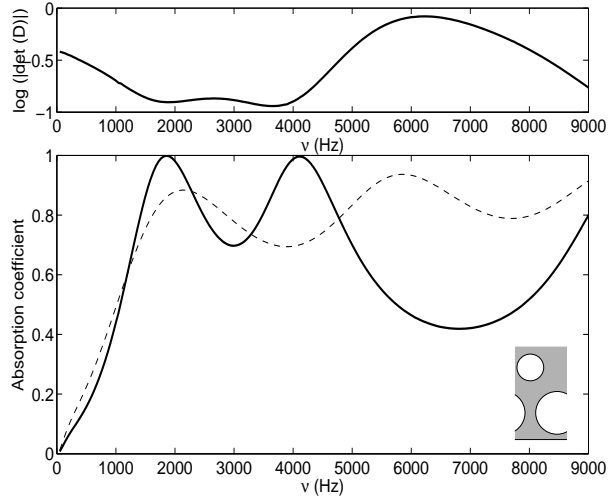}
\caption{Configuration C2 - Absorption coefficient of a $H=3.5\textrm{ cm}$ thick porous sheet of Fireflex backed by a rigid plate ($---$) without inclusion embedded and (---) with a $R^{1}=75\textrm{ mm}$ radius circular cylinder and a $R^{2}=5\textrm{ mm}$ radius circular cylinder embedded with $d=2\textrm{ cm}$.}
\label{section4s3fig1}
\end{figure}

Two absorption peaks close to unity were found for a $H=3.5\textrm{ cm}$ thick porous sheet, when a second circular cylinder of radius $R^{(2)}=5\textrm{ mm}$ is added to the configuration C1, figure \ref{section4s3fig1}. The center of this cylinder is such that $r_1^{2}=d=2\textrm{ cm}$ and $\theta_{1}^{2}=\pi/3$. The configuration C2 was derived from a triangular lattice by reducing the radius of the upper cylinder to decrease the structure thickness. The first absorption peak stands around $\nu_t^{(1)}\approx1850\textrm{ Hz}$ just below the first quarter-wavelength resonance frequency of the $H=3.5\textrm{ cm}$ thick porous sheet, and the second stands around $\nu_t^{(2)}\approx 4120\textrm{ Hz}$. These two peaks correspond to minimum of $|\textrm{det}\left(D\right)|$ and can therefore be explained by excitation of trapped modes, shifted in the complex plan.

When the inclusions are inversely placed, i.e. the center of the first inclusion is $(1\textrm{ cm},2.5\textrm{ cm})$ and $r^{(12)}=d=2\textrm{ cm}$ and $\theta^{12}=-\pi/3$, results are not identical and no quasi-total absorption peak is encoutered. This means that the configuration are not reversible.

The procedure was ran a second time with the addition of a third inclusion of radius $R^{(3)}=2.5 \textrm{ mm}$ to the configuration C2, $r_{2}^{3}=d=2\textrm{ cm}$ and $\theta_{2}^{3}=2\pi/3$. Three quasi-total absorption peaks were encoutered around $\nu_t^{(1)}\approx 1500\textrm{ Hz}$, $\nu_t^{(2)}\approx3300\textrm{ Hz}$, and $\nu_t^{(3)}\approx 5000\textrm{ Hz}$. This phenomenon was already encountered in\cite{utsunomiya1999}, where $N$ trapped modes were found when $N$ cylinders were place across a wave tank. Nevertheless, this configuration imposes use of $H=5\textrm{ cm}$ thick plate and the absorption gain was considered unsignificant over the whole frequency range considered.

The addition of inclusions that imposes increasing the thickness of the structure is rapidly of non practical use because of the large absorption of the porous layer itself.

Based on the fact that the frequency $\nu_t^{(1)}$ decreases when $x_2^{(1)}$ increases, several attempts were followed to construct a porous sheet with a unit cell composed of varying $x_2^{(j)}$ central coordinate circular cylinder arranged in a kind of garland. For example, the absorption coefficient of a $8$ circular cylinders per unit cell embedded in a $2\textrm{ cm}$ thick porous sheet was studied. The radius of the $8$ cylinders was $R^{(j)}=75\textrm{ mm}$ and the projection of the center-to-center distance between two adjacent cylinders $x_1^{(j,j+1)}$ was $2\textrm{ cm}$. The $x_2^{(j)}$ were chosen such that $x_2^{(1)}=1.1\textrm{ cm}$, $x_2^{(2)}=1.05\textrm{ cm}$, $x_2^{(3)}=1\textrm{ cm}$, $x_2^{(4)}=0.95\textrm{ cm}$, $x_2^{(5)}=0.9\textrm{ cm}$, $x_2^{(6)}=0.95\textrm{ cm}$, $x_2^{(7)}=1\textrm{ cm}$, and $x_2^{(8)}=1.05\textrm{ cm}$. The absorption peak at $\nu_{t}$ was no more total and no particular increase of its width was noticed. This means that the periodicity has a large influence on the results and that the phenomenon can not be simply explained by trapped modes but rather by complex embedded Rayleigh-Bloch waves. A similar procedure was followed by decreasing the radius and the center-to-center distance, the cylinders being aligned, without particular effects on the absorption.  
\section{Conclusion}
The influence of  embedding periodic circular inclusions on the absorption of a porous sheet attached on a rigid plate was studied theoretically and numerically. In  addition to the absorption features related to excitation of modified mode of the plate and to Bragg interference, it is shown that the structure, in case of one array of cylinders embedded in a porous sheet, whose thickness and parameters, mainly the flow resistivity, are correctly chosen, possesses a quasi-total absorption peak below the quarter-wavelength resonance frequency. This particular feature enable the design of small dimension absorption packages and was explained by complex trapped mode excitation and by increase of the pressure gradient inside the layer. This quasi-total absorption peak was validated by use of Finite Element method, thus validating the described method and results.

In case of more than one circular cylinder per spatial period, it was found that $N^{c}$ quasi-total absorption peak can be obtained for particular arrangement along the porous thickness, i.e. close to triangular lattice. Nevertheless, this rapidly leads to a large thickness of the stucture and the embedding of the additional inclusions become useless. Garland arrangment were also tested without particular effect, or at least without as spectacular effect as the one already observed for one inclusion per spatial period.

The method offers an alternative to multi-layering and double porosity materials for the design of sound absoprtion packages.
\appendix
\section{Modal analysis of the configuration}\label{app1}
\begin{figure}
\includegraphics[width=8.0cm]{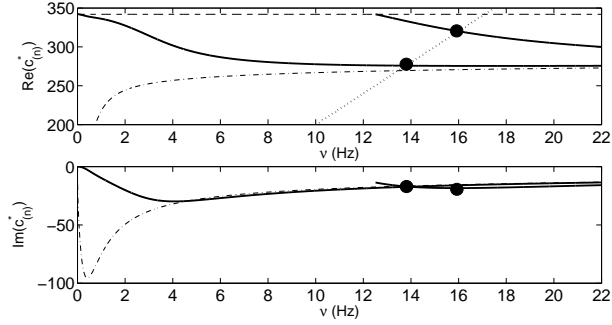}
\caption{Real and imaginary part of the root of the dispersion relation in absence of inclusions $c_{(n)}^\star$, $n=1,2$. Real part of the modified mode of the backed layer $c_{(n,q)}^\star$, $n=1,2$, $q=1,2$, for $d=2\textrm{ cm}$ are pointed out by dot.}
\label{app1fig2}
\end{figure}

The modes of the configuration without inclusions embedded (i.e. a rigid porous layer backed with a planar rigid wall), whose dispersion relation is
\begin{equation}
D^i=\alpha^{[0]i}\cos\left(k_{2}^{[1]i} H\right)-\mbox{i}\alpha^{[1]i}\sin\left(k_{2}^{[1]i} H\right)=0~,
\label{section4e1}
\end{equation}
cannot be excited by a plane incident wave initially traveling in the air medium \cite{groby_jasa2010}. Effectively, figure \ref{app1fig2} depicts the real and the imaginary parts of the roots $c_{(n)}^{\star}(\omega)=\omega/k_{1,(n)}^{\star}(\omega)$ of Eq. (\ref{section4e1}), as calculated for a $H=2\mbox{ cm}$ thick porous layer, whose acoustical characteristics are those used in section \ref{section4}. Under the rigid frame assumption and for frequencies higher than the Biot frequency (and lower than the diffusion limit), a porous material can be considered as a modified fluid, its associated dissipation being considered as a perturbation of a fluid. For Eq.(\ref{section4e1}) to be true without dissipation, $k_2^{[0]}$ should be purely imaginary while $k_2^{[1]}$ should be purely real. Under the previous assumptions, this implies that $\mbox{Re}\left(c_{(n)}^{\star}\right)$ should stand in $[\mbox{Re}\left(c^{[1]}\right),c^{[0]}]$, i.e. $|k_1^i|$ should stand in $[k^{[0]};\mbox{Re}\left(k^{[1]}\right)]$. Or for a plane incident wave initially propagating in the air medium $|k_1^{i}|$ is always smaller than $k^{[0]}$. It is also necessary to note that in the diffusion regime, i.e. for frequencies largely below the Biot Frequency, any mode exists. This fact constitutes the major difference when compared with a traditional fluid. Effectively, largely below the Biot frequency, $k^{[1]}$ is purely imaginary. This implies that $k_2^{[1]}$ is also purely imaginary whatever the value of $k_1^i$ and that $D^i$ never vanishes.

When inclusions are periodically  embedded in the porous sheet, the dispersion relation of the modes of the configuration is $\mbox{det}\left(\mathbf{I}-\mathbf{V}\left(\mathbf{S}+\mathbf{Q}\right)\right)=0$. The roots of the latter dispersion relation are difficult to determine because of the complex nature of the matrix $\mathbf{I}-\mathbf{V}\left(\mathbf{S}+\mathbf{Q}\right)$.
Here, we focuse on the case of only one grating, i.e. $N^{c}=1$ in order to emphasis the excitation of the modified mode of the backed layer (MMBL), \cite{groby_jasa2010}. Proceeding as in\cite{groby_jasa2009}, an iterative scheme can be employed to solved (\ref{section3ss2e5}). The equation is re-written in the form $(1-V_L \mathcal{M}_{LL})B_L=V_L \mathcal{F}_{L}+V_L\sum_{l\in\mathbb{Z}}\mathcal{M}_{Ll} B_l(1-\delta_{Ll})$. The iterative scheme reads as
\begin{equation}
\left\{
\begin{array}{l}
\displaystyle B_L^{\{0\}}=V_L \mathcal{F}_{L}/\left( 1-V_L \mathcal{M}_{LL} \right)\\[8pt]
\displaystyle B_L^{\{n+1\}}\!=\!\!\left(\!V_L\!\sum_{l\in\mathbb{Z}}\!\mathcal{M}_{Ll} B_l^{\{n\}}\!\left(1-\delta{Ll}\right)\!+\!V_L \mathcal{F}_{L}\!\right)\\[8pt]
/\left( 1-V_L \mathcal{M}_{LL} \right)
\end{array}
\right.,
\label{app1e1}
\end{equation}
from which it becomes apparent that the solution $B_{L}^{\{n\}}$, to any order of approximation, is expressed as a fraction, the denominator of which not depending on the order of approximation can become small for certain couples $(k_{1q},\omega)$, so as to make $B_{L}^{\{n\}}$, and possibly the field large.

When this happens, a natural mode of the configuration, comprising the inclusions and the plate, is excited, this taking the form of a resonance with respect to $B_{L}^{\{n\}}$, i.e., with respect to a plane wave component of the field in the plate relative to the inclusions. As $B_{L}^{\{n\}}$ is related to $f_p$, $g_p$ and $R_p$, the structural resonance manifests itself for the same $(k_{1q},\omega)$, in the fields of the plate and in the air.

The approximate dispersion relation
\begin{equation}
\mathcal{D}_L=1-V_L\left(S_0+\sum_{q\in\mathbb{Z}}Q_{LLq} \right)=0~,
\label{app1e2}
\end{equation}
is the sum of a term linked to the grating embodied in $V_L S_0$ with a term linked to the plate embodied in $V_L\sum_{q\in\mathbb{Z}}Q_{LLq}$, whose expressions are give in (\ref{section3ss2e2}).

This can be interpreted as a perturbation of the dispersion relation of the gratings by the presence of the plate. The zeroth order lattice sum can be re-written\cite{linton} in the form $\sum_{q\in\mathbb{Z}}2/dk_{2q}^{[1]}$ (additional constants are neglected). Introducing this expression into (\ref{app1e2}) gives
\begin{equation}
\mathcal{D}_L=1-V_L\sum_{q\in\mathbb{Z}}\frac{2\mathcal{N}_{Lq}}{dk_{2q}^{[1]} D_q}=0~,
\label{app1e3}
\end{equation}
with
\begin{multline}
\mathcal{N}_{Lq}= \alpha_q^{[1]}\cos\left(k_{2q}^{[1]}H \right)-\mbox{i}\alpha_q^{[0]}\sin\left(k_{2q}^{[1]}H \right)+\alpha_q^{[1]} \cos\left(k_{2q}^{[1]} \left(2x_2^{(1)}-H \right)-2L\theta_q \right)\\[8pt]+\mbox{i}\alpha_q^{[0]} \sin\left(k_{2q}^{[1]} \left(2x_2^{(1)}-H \right)-2L\theta_q \right)~.
\label{app1e3b}
\end{multline}

It is then convenient, for the clarity of the explanations, to consider i) $M^{[1]}$ to be a non-dissipative medium (a perfect fluid) and ii) the low frequency approximation of $V_L$, valid when $k^{[1]} R^{(1)}<\!<1$. The latter hypothesis ensures that the $V_L$ reduces to $V_l\approx (-1)^{l}\pi (k^{[1]}R^{(1)})^2/\textrm{i}4 +O\left((k^{[1]}R^{(1)})^2\right)$, $l=-1,0,1$. Equation (\ref{app1e3}) then reduces to
\begin{equation}
\mathcal{D}_l\approx 1-\sum_{q\in\mathbb{Z}}\frac{(-1)^{l}(k^{[1]}R^{(1)})^2}{2\textrm{i} dk_{2q}^{[1]} D_q/\mathcal{N}_{lq}}=0~,l=-1,0,1~.
\label{app1e4}
\end{equation}
By referring to Cutler mode \cite{cutler}, but also to the modal analysis carried out in \cite{groby_jasa2009}, the latter dispersion relation is satisfied (in the non-dissipative case) when the denominator of Eq. (\ref{app1e4}) is purely imaginary and vanishes. These conditions are achieved when $|k_{1q}|\in[k^{[0]},\mbox{Re}\left(k^{[1]}\right)]$ and when either $D_q=0$ or $\alpha_q^{[1]}=0$ (i.e. $k_{2q}^{[1]}=0$), which respectively corresponds to modified modes of the backed-layer (MMBL) and to modes of the grating (MG). Both of them are determined by the intersection of $c_{1q}=\omega/k_{1q}$ respectively with $c_{(n)}^{\star}(\omega)$ as calculated for the backed-layer and with $\mbox{Re}\left(c^{[1]}\right)$. The MMBL are pointed out by the dots on figure \ref{app1fig2}. The associated attenuation of each mode can then be determined by the values of $\mbox{Im}\left(c_{(n)}^\star\right)$ and $\mbox{Im}\left(c^{[1]}\right)$ at the frequencies at which the modes are excited. The attenuation associated with MG is also higher than the one associated with MMBL for all frequencies. Moreover, MG corresponds to the highest boundary of $|k_{1q}|$ for Eq.(\ref{app1e4}) to be true. This implies that MG should be difficult to excit. The latter type of mode can only be poorly excited by a plane incident wave, particularly at low frequencies.

\end{document}